\documentclass[12pt]{emulateapj}

\shorttitle{Fraction of RV-stable stars in 2001-2003 GGSS sample}
\shortauthors{Bizyaev \& Smith}

\begin{document}

\title{Fraction of the radial velocity stable stars in the early 
observations of the Grid Giant Star Survey}

%\author{Dmitry Bizyaev\altaffilmark{1,2}
%\and
%Verne V. Smith\altaffilmark{1,3}}
%\email{dmbiz@noao.edu, vsmith@noao.edu}

\author{Dmitry Bizyaev$^{1,2}$
and
Verne V. Smith$^{1,3}$}
\email{dmbiz@noao.edu, vsmith@noao.edu}

\altaffiltext{1}{National Optical Astronomy Observatory, Tucson, AZ, 85719}
\altaffiltext{2}{Sternberg Astronomical Institute, Moscow, 119899, Russia}
\altaffiltext{3}{McDonald Observatory, University of Texas, Austin, TX 78712}

\begin{abstract}

The GGSS is a partially-filled, all-sky survey to identify K-giant stars
with low level of RV-variability.  We study histograms of the radial
velocity (RV) variability obtained in the early phase of the Grid Giant Star
Survey (GGSS, Bizyaev et al., 2006). This part of the survey has been
conducted with a very limited nubmer of observations per star, and rough
accuracy. We apply the Monte-Carlo simulations to infer a fraction of the
RV-stable stars in the sample. Our optimistic estimate is that 20\% of all
considered K-giants have RV-variability under 30 m\,s$^{-1}$. Different
assumptions of intrinsic RV-variability for our stars give 12 -- 20 \% of
RV-stable K-giants in the studied sample.

\end{abstract}


\section{Introduction}

The future space interferometric missions have to be supplied with a
reference frame of objects uniformly spaced in the sky, bright enough, and
indicating no radial velocity (RV hereafter) oscillations at a high level of
accuracy (tens m/s). A large sample of relatively faint
K-giants (V = 9.5 - 13.5 mag) that may serve as candidates
to the reference sources for the {\it Space Interferometry Mission (SIM
Planet Quest)} has been studied in 2001-2004 by \citet{paper1} in the frames
of a preparatory survey GGSS \citep{patterson01}. As it was found by
\citet{paper1}, only about 1/3 of all candidates reveal the RV-variation
within 100 m\,s$^{-1}$ level, when observed with one year cadence. However, most of
the objects in that sample were observed only 2-3 times.  Therefore, the
real, or intrinsic variability of K-giants can not be estimated directly
because of the limited number of RV-observations per star, and also because of
non-zero accuracy of individual RV estimations. A more accurate study of the
RV-variation of bright K-giants which was carried out by \citet{frink01},
suggests a higher fraction of RV-stable stars among K-giants. On the other
hand, well representative number of observations per star (from 11 to 28)
reveals 34 of 179 RV-stable bright K-giants in the study by
\citet{Hekker06}. Studies by \citet{frink01} and \citet{Hekker06} 
suggest a correlation between the RV variability and color (B-V). 
Contrary to them, this correlation was not found by \citet{paper1}, that
can be explained by worse accuracy of RV estimates in the latter case.

In this paper we undertake a study of histograms of the RV-variability 
published for the GGSS survey by \citet{paper1}.  Monte-Carlo
simulations that take into account the limited number of observations per
star and rough level of the individual RV-estimations are performed.  The main
idea of the simulations is that the intrinsic (i.e. real and unknown)
distribution of the stellar RV-variability is the same for both Northern (N)
and Southern (S) subsamples \citep[see ][for detail]{paper1}. Since these
two subsamples were observed with different instruments, and since different
techniques of the RV estimation were applied, the observational
uncertainties of the RV estimations in our simulations may be
different for the subsamples. The subsamples differ from each other also 
in the number of observations per star.

Resulting model histograms of the RV-variability 
\citep[or sigma, as it was designated by][]{paper1} depend on both
intrinsic RV-variability and factors introduced by observations.
The published distributions of the RV-variability may be 
approximated by the model histograms by adjustment of model parameters.

\section{Modeling of the intrinsic sigma distributions}

We use the following basic assumptions to model the observational
histograms presented by \citet{paper1}. 

(1) Distribution
of the intrinsic RV-variability (or intrinsic sigma, hereafter)
is assumed to be the same for both N- and 
S-subsamples in our modeling. The shape of the intrinsic sigma
distribution to be chosen later in this section.
We note that distribution of the RV-variability
is unknown for any limited number of observations.

(2) Obsevations introduce uncertainties into the resulted RVs.
We add normally distributed uncertainties to each individual RV.

(3) We consider the realistic sample sizes: $N_{stars}$ = 148 
and 341 for N- and S-subsamples, rspectively. The realistic numbers of
observations per star $N_{obs}$ = 2 and 3 for the N- and S-subsamples, 
respectively, are assumed in the simulations.

For a certain value of the intrinsic RV variability,  
we generate $N_{obs}$ normally distributed random radial velocities (mean
value of the radial velocity is zero, the standard deviation is the 
intrinsic sigma). Then we add a normally distributed random "observational" 
uncertainty to each RV. We assume the standard deviation of this 
uncertainty as a free parameter in the modeling.  
The obtained model values of RV are used to calculate  
the mean radial velocity and its
standard deviation (i.e. the model sigma) exactly in the manner done 
with the real observations \citep[see the procedure described by][]{paper1}. 
The described procedure is repeated $N_{stars}$ times.
It gives us $N_{stars}$ values of the observed RV-variability.

The described simulation is repeated 40 times with random values of RV. 
For the comparison purposes, certain models are evaluated 100 times to 
make sure 
that results are independent of the number of simulations.  
The resulting values of sigma are calculated for $N_{stars}$
simulations. The final RVs are obtained by averaging over all 40 repeated
calculations. The obtained model sigmas are sorted out 
into the same bins in the same way as in histograms obtained
from observations \citep{paper1}. 
As a result, we obtain the mean numbers of model sigma in each bin 
of N- and S- histograms.

For the comparison of the model histograms and observed RV-variability
distributions, we figure out one chi-square value $\chi^2$ which is common
for both N- and S- subsamples.  The contributions of the samples into the
$\chi^2$ are weighted proportional to the size of the samples.

%In some special cases when the intrinsic sigma supposed to 
%take different values with some probability (i.e. if sigma can take multiple 
%values, or be continuous), we evaluate the model described above for the 
%corresponding subset of sigmas, and then mix the resulting model 
%distributions with weights according to their probability.

We consider the following three assumptions on the functional form of
the intrinsic RV variability (sigma) in the modeling.

(1) All stars have the same single value of the sigma.

(2) There are two samples of stars: "RV-stable" and "RV-unstable", and
hence there are two values of the sigma. 

(3) The intrinsic RV-variability is a continuous value which 
uniformly spans a range between 0 and a certain value.

The accuracy of the RV estimations introduced by the 
observations is a free parameter in all three cases.
In the first case, an additional free model parameter is the 
intrinsic sigma.
More flexibility is implied in the second case, where
the "stable" (typically less than 50 m\,s$^{-1}$) and "unstable" (typically greater 
than 100 m\,s$^{-1}$) sigmas, and fractions of stable and unstable stars
are additional free parameters in the modeling.
In the third case, the upper limit of the continuous sigma (the lower 
one is set at 0) is the free parameter.
In the latter case, we approximate the continuous sigma by its 
100 discrete values.

%The optimal values of the model parameters mentioned above are found by
%means of minimization of the resulting single chi-square value.

\section{Results of the modeling}

The modeling shows that agreement between the model and observing
histograms of distribution of the RV-variability can be achieved
with quite wide ranges of free parameters, whereas
the fiting of the southern histogram is more difficult.
Our model \#1 does not agree well with the observations. 
The best-fit intrinsic RV-variability in the model \#1 takes
high values of the order of several hundreds m/s, and the best-fit
accuracy of both N- and S-observations is rough (several hundred m/s). 
The minimum $\chi^2$ value is about 3.5 times higher than that obtained in
the next two models.

%%Model \#2 gives better agreement with the observational histograms.

\begin{figure}   
%\plotone{f1.eps}
\includegraphics[scale=.40]{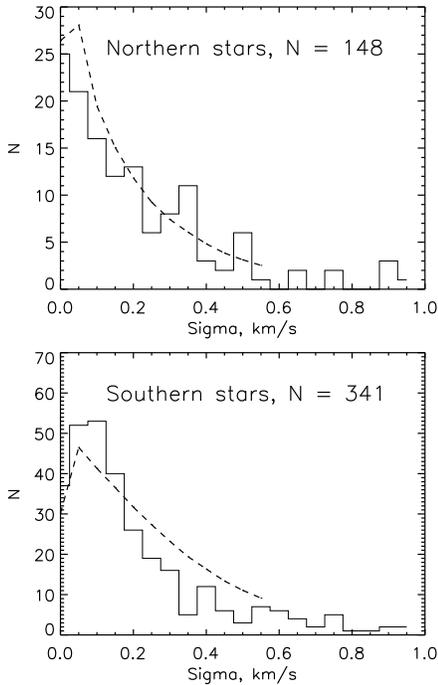}
\caption{Distributions of observed RV-variability (sigma) for
the Northers (top) and Southern (bottom) samples
are shown by the solid lines. The dashed lines connect the
points on the model distribution of the RV-variability. The latter
distribution corresponds to the best-fit solution obtained within
the model \#2 (see text).}
\label{f1}  
\end{figure}

The best-fit intrinsic sigma of the "stable" stars in the model \#2 
ranges from 10 to 30 m\,s$^{-1}$, whereas the "unstable" stars have
the sigma of 890 m\,s$^{-1}$. The best-fit accuracy of the RV estimations in 
this case is about 100 and 50 m\,s$^{-1}$ for the southern and northern 
samples, respectively. 
The best-fit fraction of the RV-stable stars is 20\% in our model \#2. 
Figure \ref{f1} shows the histogram based on observations (solid) 
and the best-fit model distribution of the RV-variability (dashed). 
A higher fraction of the RV-stable stars 
would make the central peaks of the model histograms too high
in both top and bottom panels. 
A lower fraction of the RV-unstable stars would create a
disagreement in the tails of both N- and S-histograms.

\begin{figure}   
%\plotone{f2.eps}
\includegraphics[scale=.40]{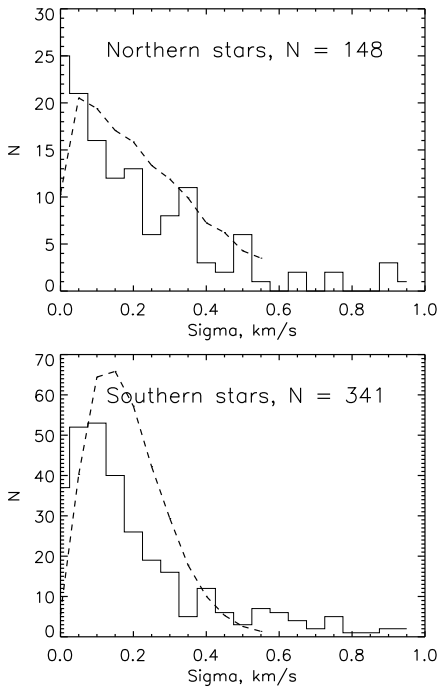}
\caption{The same as in Figure \ref{f1} for the model \#3.}
\label{f2}  
\end{figure}

Model \#3 gives the best $\chi^2$ value of all considered models. 
The best-fit observational RV-uncertainty is 50 m\,s$^{-1}$ for both N- and 
S-samples. The best-fit intrinsic sigma is distributed uniformly
between 0 and 850 m\,s$^{-1}$. Assuming 100 m\,s$^{-1}$ as the limit between 
the RV-stable and unstable stars, one can obtain 12\% as a fraction
of the RV-stable stars in our sample. Figure \ref{f2}
shows the best-fit model and observing histograms of 
the RV-variability for the model \#3.

Although formally the minimum $\chi^2$ value in the model \#3 is less 
that that in the model \#2, the difference is not significant.
At the same time, the comparison of Figures \ref{f1} 
and \ref{f2} shows that the model \#2 better matches
the shape of the observing N- and S- histograms.

The real distribution of the intrinsic RV-variability 
can have an arbitrary shape. More precise studies
of the RV-variability suggest a one-peak distribution of the intrinsic
RV-variability and a higher fraction of the RV-stable stars than that 
found in our simulations. 
In order to make the fraction of the RV-stable stars higher in our modeling,
one had to assume that the maximum of the intrinsic sigma distribution
would have been shifted toward the lowest values of the RV-variability.
On the one hand, our model \#3 is an extreme case for such
class of the intrinsic sigma distributions, and hence 12\% is the 
lower limit for the fraction of stable stars. On the other hand,
the model \#2 is another extreme case, since the uniform
distribution of the intrinsic sigma is a marginal case of 
arbitrary monotonously descending distributions.
As a result, the real fractions of the RV-stable and unstable stars
would not be that much different from 20\% and 80\%, respectively.

%If a monotonously descending distribution of
%the intrinsic sigma with the maximum between 10 and 30 m\,s$^{-1}$ is considered, 
%then the bulk of such distribution corresponds to our subsample of 
%the RV-stable stars in the model \#2. According to our modeling,
%this RV-stable subsample may consist of 20\% of all stars. 
%The "tail" of the intrinsic RV-variability distribution 
%(intrinsic sigma greater than 30 m\,s$^{-1}$) would encompass 80\% of all
%stars, and the average value of sigma will tend to be high, 

Any deviations from our assumptions about the model parameters may
change the assessed fraction of the RV-stable stars. Thus,
introduction of systematic errors into measurements of RVs 
makes the model histograms wider, allowing the fraction of
bona fide RV-stable stars to be higher in this case. The observed
RVs for the southern sample show worse agreement with our modeling 
than the northern RVs under any assumption of the balance between 
the fractions of RV-stable and unstable stars.
In the case of the southern sample, we might encounter with systematic
errors in the observing RVs.
Note that the southern sample spectra were obtained with rather low
S/N, of the order of 10 and even less \citep[see][for details]{paper1}. 
However, most of the northern stars were observed just
two times. It introduces more uncertainties into the modeling for the 
northerm subsample.

\section{Conclusions}

Our Monte-Carlo simulations of observed distributions of the RV-variability  
reveals 12 -- 20 \% of the RV-stable stars in the
2001-2004 sample of the Grid Giant-Star Survey under basic assumptions on
the intrinsic RV-variability distribution. Our optimistic estimation is
that 20 \% of all considered K-giants have the intrinsic 
RV-variability under 30 m\,s$^{-1}$.  
It is in agreement with a recent study of the RV-variability performed for
bright K-giants by \citet{Hekker06}, but less than that found by
\citet{frink01}.  As it was discussed by \citet{paper1}, constraining
the physical parameters of the considered candidates can help to increase 
the fraction of RV-stable stars, and partially explains the difference in 
the stable stars fractions reported in different papers.

The universal receipt for a radial velocity study intended to
select the RV-stable stars is to conduct
as many observations per star as possible, and to achieve the accuracy of
individual RV-estimations that is better than the required level of the
RV-stability. In this case, the RV-variability histograms obtained from 
the observations will reveal the shape of the intrinsic sigma distribution. 
It reflects the spirit of recent searches for the RV-stable stars
for prospective space interferometric missions
\citep{Hekker06}, and agrees with the ideas and design of the ongoing SIM
Planet Quest grid candidates observations where the iodine cell technique is
being applied.

\begin{acknowledgments}
The GGSS follow-up observations have been
supported by the JPL and NASA via grant 99-04-OSS-058.
\end{acknowledgments}

%\clearpage

%\begin{figure}
%%\plotone{f1.eps}
%\includegraphics[scale=.40]{f1.eps}
%\caption{Distributions of observed RV-variability (sigma) for
%the Northers (top) and Southern (bottom) samples
%are shown by the solid lines. The dashed lines connect the
%points on the model distribution of the RV-variability. The latter
%distribution corresponds to the best-fit solution obtained within 
%the model \#2 (see text).}
%\label{f1}
%\end{figure}
%
%\begin{figure}
%%\plotone{f2.eps}
%\includegraphics[scale=.40]{f2.eps}
%\caption{The same as in Figure \ref{f1} for the model \#3.}
%\label{f2}
%\end{figure}


\begin{thebibliography}{}

\bibitem[Bizyaev et al.(2006)]{paper1}
Bizyaev, D., Smith, V.V., Arenas, J., et al., 2006, \aj, 131, 1784,
Paper I

\bibitem[Hekker et al.(2006)]{Hekker06}
Hekker, S., Reffert, S., Quirrenbach, A., et al., 2006., \aap, 454,
943

\bibitem[Frink et al.(2001)]{frink01}
Frink, S., Quirrenbach, A., Fischer, D., R{\" o}ser, S., \& Schilbach, E.
2001, \pasp, 113, 173

\bibitem[Patterson et al.(2001)]{patterson01}
Patterson, R. J., Majewski, S. R., Slesnick, C. L., et al.
Small Telescope Astronomy on Global Scales, ASP Conf. Ser. v.246,
IAU Colloquium 183. Ed. by Bohdan Paczynski, Wen-Ping Chen,
and Claudia Lemme. San Francisco: Astronomical Society of the
Pacific, 2001, p.65


\end{thebibliography}
\end{document}